# Hierarchical Interdiffusion Kinetics in Nanoscale Ni/Al Multilayers

Sascha S. Riegler[1], Isabella Gallino[2], Nicolas J. Peter[3], Andrey Tarasov[1,2], Tobias Meyer[4], Jörg Schmauch[5], Christoph Pauly[6], Maximilian Frey[1], Yesenia H. Sauni Camposano[7], Heike Bartsch[7], Ruth Schwaiger[3], Peter Schaaf[7], Ralf Busch[1] and Jonas Arlt[2*]

*[1] Chair of Metallic Materials, Saarland University, Campus C6.3, 66123 Saarbrücken, Germany*

*[2] Chair of Metallic Materials, TU Berlin, Ernst-Reuter-Platz 1, 10587 Berlin, Germany*

*[3] Institute of Energy Materials and Devices (IMD-1), Forschungszentrum Jülich GmbH, 52428 Jülich, Germany*

*[4] Institute of Materials Physics, University of Goettingen, D-37077 Göttingen, Germany*

*[5] Physics Department, Saarland University, Campus D2.2, 66123 Saarbrücken, Germany*

*[6] Center for Correlative Microscopy and Tomography CoMiTo, Saarland University, Campus D3.3, 66123 Saarbrücken, Germany*

*[7] Chair Materials for Electrical Engineering and Electronics, Institute of Materials Science and Engineering, Institute of Micro- and Nanotechnologies MacroNano, TU Ilmenau, Gustav-Kirchhoff-Str. 5, 98693 Ilmenau, Germany*

---

[*] corresponding author: Jonas Arlt, jonas.arlt@tu-berlin.de

# Abstract

Reactive metallic multilayers store chemical energy that can be released rapidly through interdiffusion and intermetallic formation. Predictive control of this heat release requires distinguishing transport and phase-formation processes that occur in rapid succession. Here we combine free-standing nanoscale Ni/Al multilayers with chip-based flash calorimetry and isoconversional kinetic analysis over five orders of magnitude in heating rate. Selected reaction states are quenched and examined by scanning transmission electron microscopy. This workflow separates pre-ignition interdiffusion into two regimes and quantifies activation energies of (81 ± 24) and (168 ± 17) kJ/mol, consistent with grain-boundary and lattice diffusion of Ni in Al, respectively. Microscopy supports this assignment: no significant compositional changes are observed after the first regime, whereas the second increases the Ni content of the Al layers and produces Ni-enriched features spaced by 5–10 nm, matching the Al grain size. These results identify the Al grain-boundary network as the dominant low-barrier pathway, providing rapid transport across the Al layers and priming lattice-mediated mixing and intermetallic phase formation. More broadly, the workflow links calorimetric signatures to pathway-specific kinetics and transient microstructures, enabling direct assessment of how microstructural design redirects coupled transport and reaction pathways in reactive multilayers and other materials driven far from equilibrium.

**Keywords:** Ni/Al reactive multilayers, Interdiffusion, Isoconversional kinetics, Grain boundary diffusion, Fast differential scanning calorimetry (FDSC), Four-dimensional scanning transmission electron microscopy (4D-STEM)

# Graphical Abstract

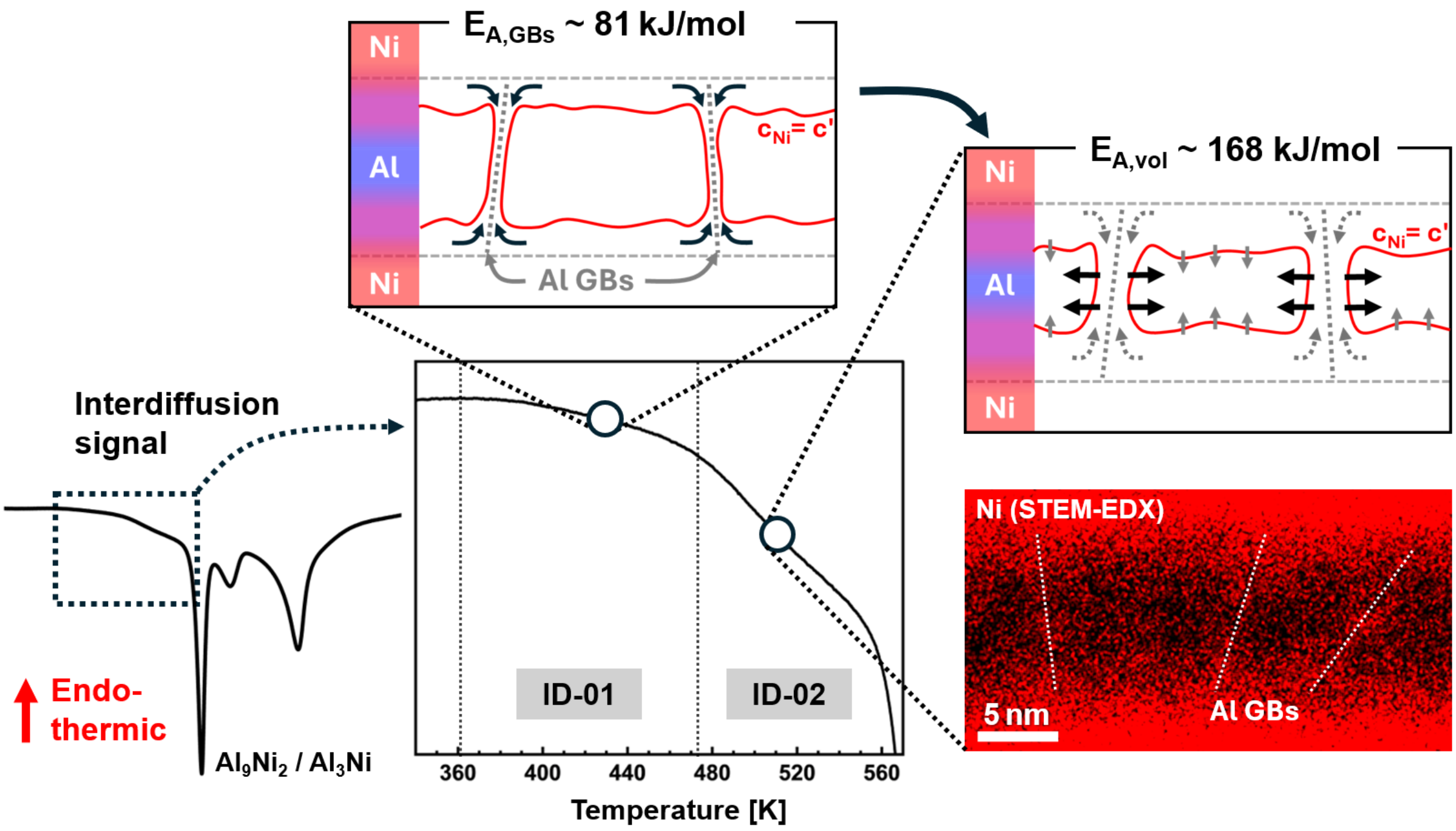

# 1. Introduction

Reactive metallic multilayer (RML) thin films consist of alternating nanometer-scale reactant layers that store large amounts of chemical energy. This energy can be released on demand by igniting a self-propagating high-temperature synthesis (SHS) reaction.[1,2] Following a local electrical, thermal, or mechanical stimulus, solid-state interdiffusion initiates heat release, while subsequent melting and intermetallic formation accelerate mixing and sustain reaction propagation through the multilayer film at velocities of 0.1–100 m/s.[2–7] This rapid, localized heat release enables applications ranging from reactive joining[8,9] to micropropulsion units,[10] synthesis of high-temperature compounds,[11] and thermal-battery heat sources,[12,13] while also offering potential for biohazard neutralization.[14,15] Importantly, these use cases impose distinct requirements on the heat-release profile: reactive joining benefits from rapid heat release and high peak temperatures, whereas thermal batteries or biohazard neutralization may require a more gradual or otherwise tailored thermal profile.[16] Consequently, predictive control over the kinetic processes governing heat release is essential for tailoring RML performance to application-specific requirements.

These processes are controlled by RML morphology and microstructure. The primary geometric parameter is the bilayer periodicity, $\Lambda$, which sets the characteristic transport distance and strongly influences ignition and SHS propagation behavior.[16] The layer microstructure, including grain size and volumetric defects, further modifies the available diffusion pathways and can affect reaction kinetics and thermal response on a comparable scale.[17–22] In Ni/Al multilayers, for instance, fast transport along Al grain boundaries accelerates solid-state interdiffusion in the pre-ignition regime, thereby affecting phase formation, local melting, and the resulting temperature profile.[17,18,21,23,24] Accordingly, microstructure engineering offers a promising route for tailoring RML performance, a potential increasingly explored through computational modeling.[21–23,25–27] However, predictive models often rely on effective, pathway-averaged diffusion coefficients and activation energies, which cannot account for sequential transport mechanisms along the reaction path. Experimental access to pathway-specific kinetic parameters under ignition and SHS-relevant heating conditions is therefore essential for predictive RML design.[28,29]

Chip-based nanocalorimetry offers a direct route to access such pathway-specific kinetic parameters. Ignition in RMLs occurs during rapid solid-state heating, typically on the order of $10^3$–$10^4$ K/s,[30] while subsequent SHS propagation subjects the multilayer to effective heating rates exceeding $10^6$ K/s.[30,31] Whereas conventional calorimetric studies of RMLs have been restricted to heating rates below a few K/s,[32–35] nanocalorimetry enables heating rates up to $10^4$–$10^6$ K/s, thereby accessing ignition-relevant conditions and extending into the regime associated with SHS propagation.[36,37] These fast rates are also essential for kinetic separation: under slow scanning conditions, closely successive transport and phase-formation processes can overlap or proceed faster than they can be resolved calorimetrically, causing extracted activation energies to reflect effective averages over multiple contributions. Rapid quenching within the FDSC further enables selected intermediate states to be preserved for ex-situ microstructural analysis, a capability difficult to achieve in bulk-scale samples because of their larger thermal mass.

In Ni/Al multilayers, nanocalorimetry has substantially refined the understanding of pre-ignition interdiffusion and intermetallic phase formation.[38–43] These studies showed that solid-state Ni transport into Al is central to ignition and supported Al grain boundaries as fast diffusion pathways initiating phase formation. However, resolving how the apparent activation energy evolves with reaction progress, as successive transport pathways and phase-formation processes become active, has so far been demonstrated only in isolated cases[44] and has generally not been linked to direct nanoscale validation of arrested transient states. A central limitation is the restricted throughput of custom-built nanocalorimetry approaches, in which reactive multilayers are deposited directly onto the calorimetric sensor, making experiments effectively single-use.[41,42,45] This architecture hinders systematic exploration of kinetic regimes and limits direct correlation of calorimetric signatures, kinetic parameters, and microstructural evolution across transient states.

Here, we introduce a transferable workflow for resolving rapid solid-state transformations in nanoscale multilayer systems. Free-standing reactive multilayer films are combined with a chip-based calorimetry platform (Mettler-Toledo FDSC2+), enabling high-throughput kinetic mapping over heating rates spanning conventional calorimetry to SHS ignition-relevant conditions. Targeted quenching followed by advanced electron microscopy further links selected heat-flow signatures directly to microstructural evolution along the transformation path. We demonstrate the capabilities of this approach by focusing on solid-state interdiffusion in nanoscale Ni/Al RMLs. The resulting data resolve the pre-ignition interdiffusion regime in unprecedented detail. The workflow not only separates closely successive transport processes, but also quantifies their distinct kinetic barriers. Beyond these mechanistic insights, this work establishes a basis for probing how microstructure engineering modifies kinetic barriers and transformation pathways in reactive multilayers and other nanostructured materials.

# 2. Results

## 2.1 Microstructure of the as-prepared multilayers

The microstructure of the as-prepared Ni/Al multilayers was examined by four-dimensional scanning transmission electron microscopy (4D-STEM, **Figure 1**). Virtual bright-field imaging resolves the periodic stacking of ~12 nm Al and ~8 nm Ni layers (**Figure 1a**), consistent with nominal deposition thicknesses. Both constituents exhibit columnar grain morphologies, with fine Ni grains (~2–5 nm) and substantially larger Al grains (~5–10 nm and locally several tens of nanometers wide; **Figure 1b,c**). Furthermore, both layers predominantly exhibit a {111} out-of-plane texture, as revealed by the virtual dark-field image generated from the {222} reflection (**Figure 1b**).

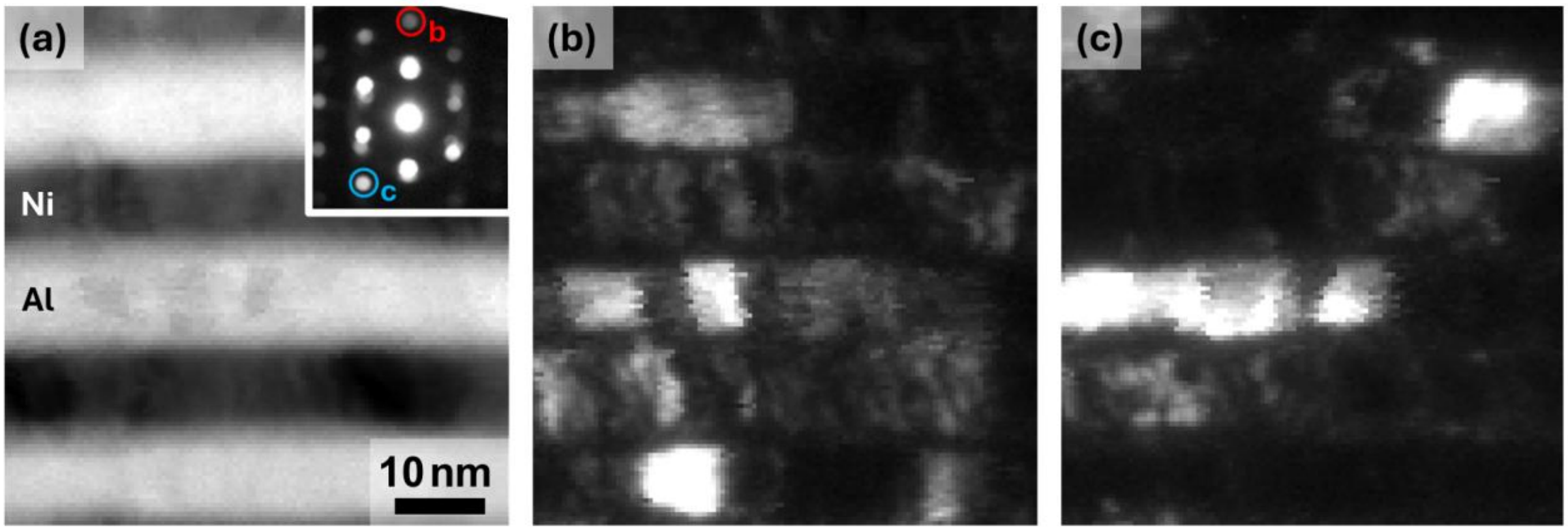

**Figure 1: 4D-STEM analysis of the as-prepared Ni/Al multilayers.** (a) Virtual bright-field image and summed diffraction pattern generated from an Al layer. (b,c) Virtual dark-field images generated from the {222} (b) and {220} (c) Bragg reflections indicated in (a), with residual Ni-layer contrast arising from partial inclusion of nearby Ni reflections. The {222} reflection probes crystallites with a {111} out-of-plane orientation. Horizontal striations in the images arise from scanning artefacts.

The interfacial sharpness of the as-prepared multilayers was evaluated using high-angle annular darkfield (HAADF) STEM. HAADF imaging reveals a pronounced interface roughness, with variations of several nanometers over lateral distances of ~30 nm (**Figure 2a,b**). The Ni layers exhibit largely uniform contrast with faint columnar features originating from the nanoscale grain structure, whereas the Al layers appear speckled with subtle intensity variations.

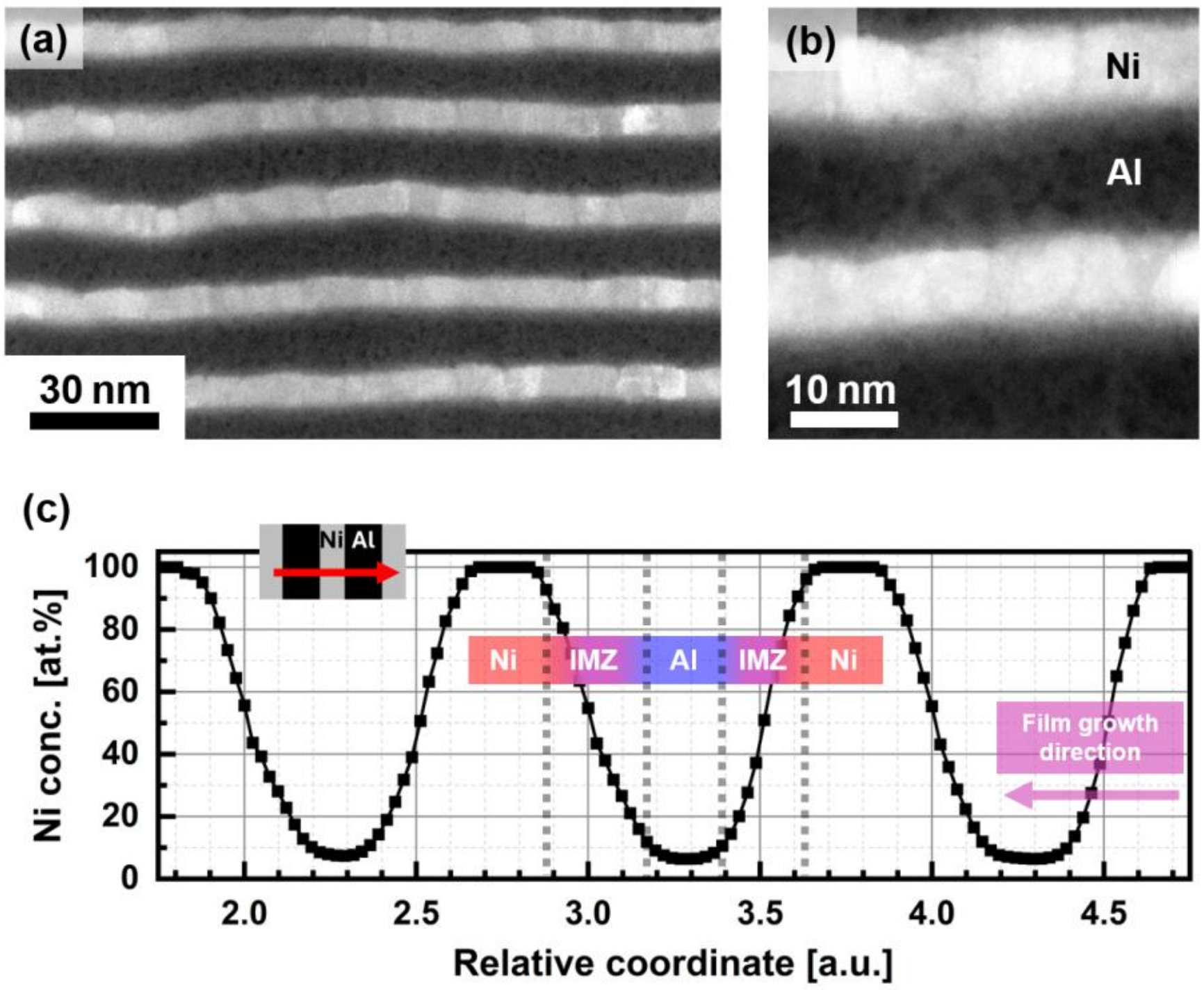

**Figure 2: Microstructure and composition of the as-prepared Ni/Al multilayers.** (a) Low-magnification STEM-HAADF overview of the multilayer stack. (b) Higher-magnification STEM-HAADF image highlighting the interfacial morphology and contrast. (c) STEM-EELS-derived Ni concentration profile across the multilayer stack.

Electron energy loss spectroscopy (EELS) was used to assess the composition of the as-prepared multilayers (**Figure 2c**). To account for the interfacial roughness and locally varying layer thickness, a relative coordinate was used in which the points of 50 at.% concentrations are mapped to fixed positions as detailed in **Supplementary Note S1**. The profiling reveals extended intermixing zones (IMZs) between the Ni and Al layers, spanning up to 7 nm in some cases (**Figure SI 2**). Notably, there is a pronounced asymmetry: Ni grown on Al produces larger IMZs than Al grown on Ni. In addition, the Al layers show pronounced through-thickness enrichment in Ni, with concentrations in the layer centers remaining above ~6 at.%, whereas the centers of the Ni layers remain largely free of Al. While these observations demonstrate substantial Ni enrichment of the Al layers, the measured IMZ widths do not solely reflect chemical intermixing. Instead, they contain contributions from genuine interdiffusion, projection-induced overlap caused by interface roughness, and beam broadening due to electron scattering within the TEM lamella. Consequently, the experimentally observed ~7 nm IMZ width should be regarded as an upper bound for the actual extent of chemical intermixing.

The observed premixed state is consistent with previous reports on sputter-deposited Ni/Al multilayers, where intermixed interface regions and substantial Ni enrichment of the Al layers have been reported,[46–52] particularly for small bilayer periodicities ($\Lambda < 50$ nm).[48,51] A similar asymmetry in IMZ width has also been observed.[50,53] Such asymmetric intermixing appears to be a broader feature of sputter-deposited Al/transition-metal systems,[53] although its origin is not yet fully understood and likely involves both thermally activated interdiffusion and ballistic mixing during film growth (see **Supplementary Note S2**).

## 2.2 Calorimetric signatures and phase evolution

The early interdiffusion regime in equimolar Ni/Al multilayers was investigated using conventional DSC and flash DSC (FDSC) over a heating-rate range spanning more than five orders of magnitude, from 0.1 to $10^4$ K/s. Across all investigated heating rates, the heat-flow curves exhibit a characteristic sequence of sharp exothermic peaks preceded by a broad, shallow exothermic shoulder (**Figure 3**). The distinct peaks (A–C) are associated with the sequential formation of Al–Ni intermetallic phases,[31,38,54–56] including, $Al_9Ni_2$, $Al_3Ni$, $Al_3Ni_2$, and B2-AlNi and are analyzed in detail in a companion publication. In contrast, the preceding exothermic shoulder is commonly attributed to interdiffusion prior to intermetallic formation.[44,46,57–59] In the present FDSC data, this shoulder separates into two distinct interdiffusion (ID) intervals, denoted ID-01 and ID-02, which are distinguished by a clear change in slope of the heat-flow signal.

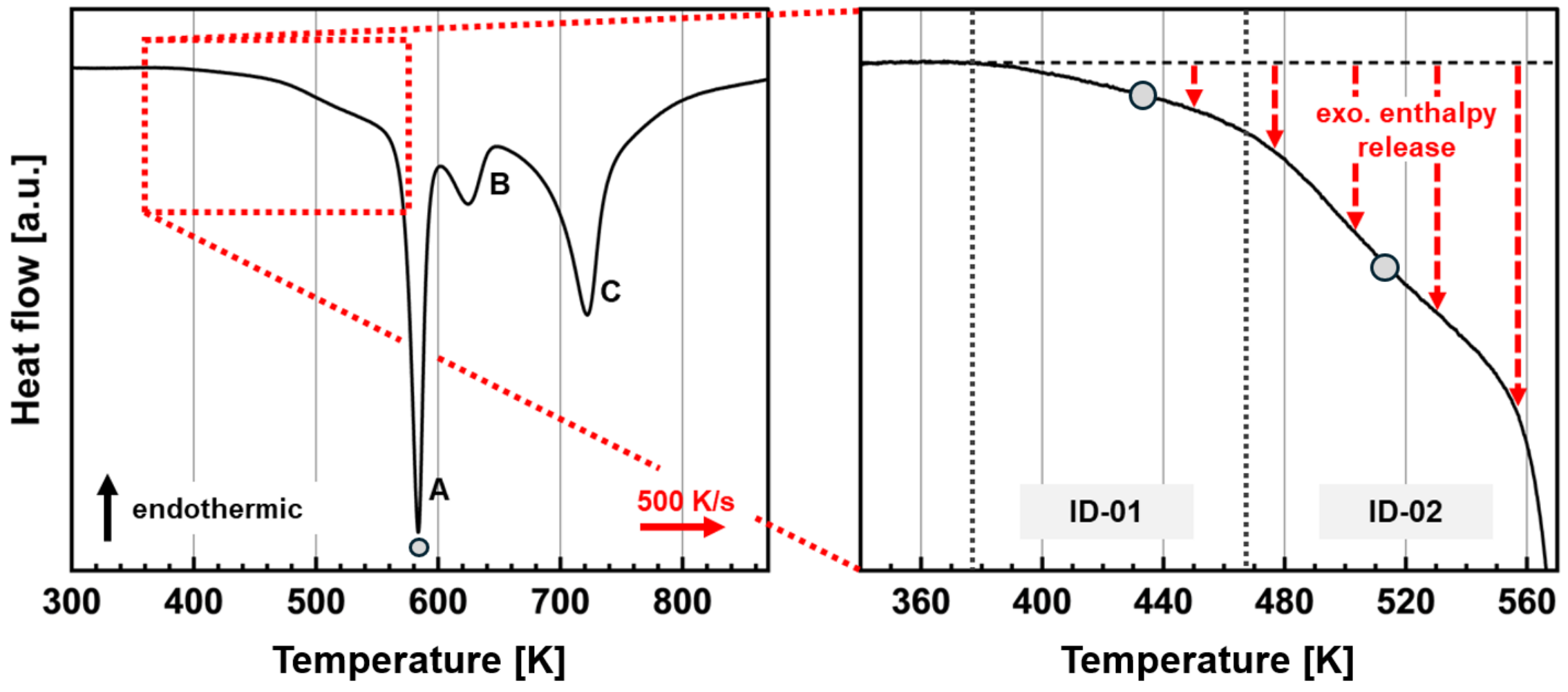

**Figure 3: Representative heat flow curve of an equimolar Ni/Al multilayer ($\Lambda = 20$ nm) recorded at 500 K/s (FDSC, UFH1 sensor).** Exothermic peaks A–C mark the main stages of Ni–Al intermetallic formation. The enlarged view highlights the preceding interdiffusion regime, which resolves into two intervals (ID-01 and ID-02). Grey circles indicate reaction stages selected for subsequent microstructural analysis (ID-01, ID-02, and Peak A).

The separation between ID-01 and ID-02 is most clearly resolved at heating rates above 10 K/s using the UFH1 sensor, whereas it is only occasionally discernible in measurements performed with the UFS1 sensor at comparable rates (**Figure 4a,b**). At heating rates above approximately 2500 K/s, the distinction again becomes less pronounced. In this regime, thinner multilayer specimens (1 µm instead of 5 µm) were required to suppress ignition of a self-propagating SHS reaction. Notably, the onset temperatures of both ID intervals shift systematically to higher temperatures with increasing heating rate, similar to the subsequent exothermic peaks associated with intermetallic phase formation (**Figure 4a,c**). This heating-rate dependence indicates that both ID-01 and ID-02 result from thermally activated processes. The restricted range over which the two intervals can be resolved likely reflects their rapid progression and comparatively low enthalpy release. Resolution of the separation is favored by the higher sensitivity of the UFH1 sensor and the increased heat-flow intensity at faster heating rates, which facilitate detection of the comparatively weak ID-01 contribution. At the highest rates, however, the reduced sample mass counteracts this signal enhancement.

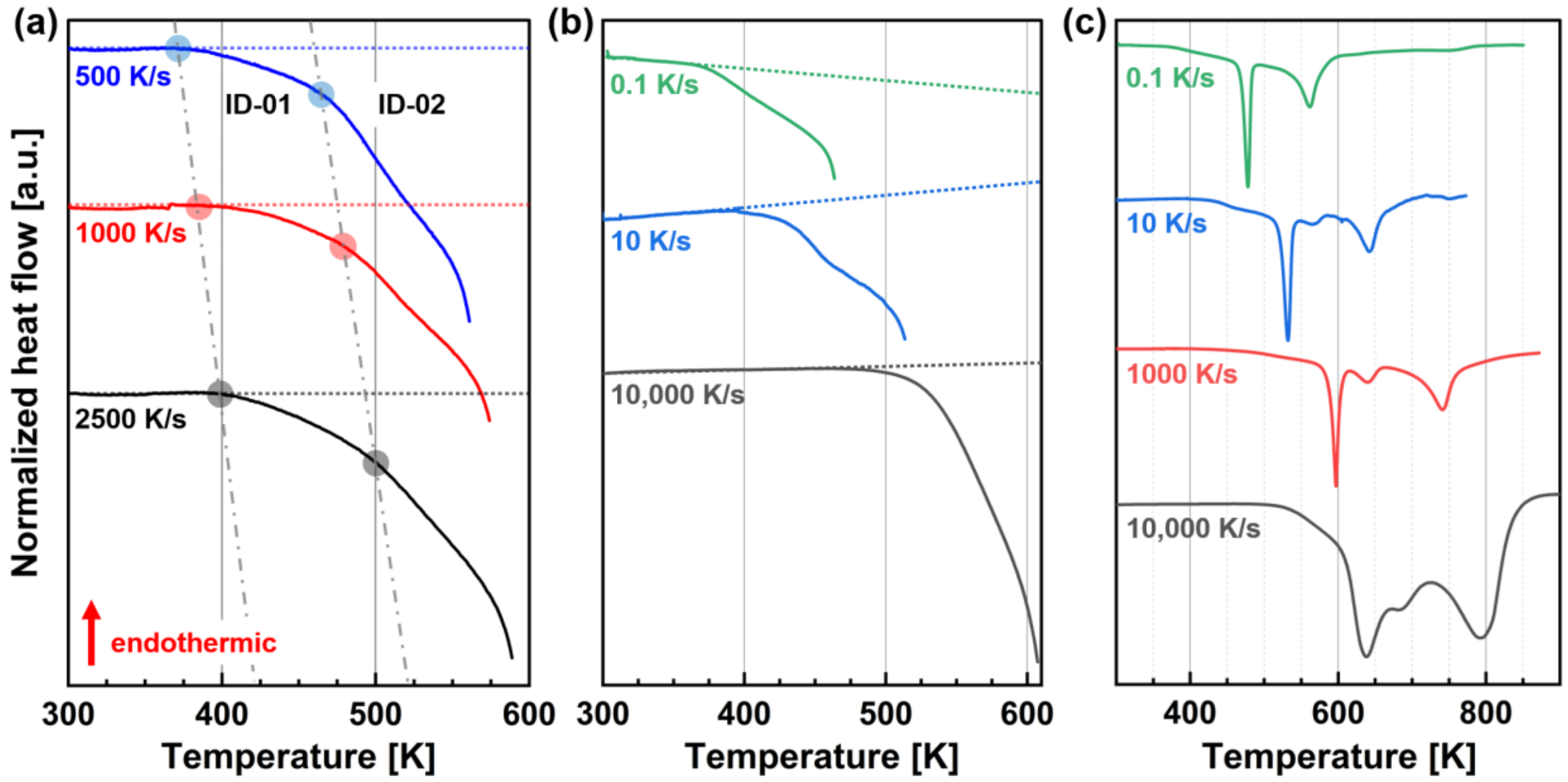

**Figure 4: Heating-rate dependence of the heat-flow response of equimolar Ni/Al multilayers ($\Lambda$ = 20 nm).** (a) Enlarged view of the interdiffusion regime measured with the UFH1 sensor at 500, 1000, and 2500 K/s. Circles mark the onsets of ID-01 and ID-02; dash-dotted lines serve as guides to the eye. (b) Enlarged interdiffusion-region curves recorded at 0.1 K/s by conventional DSC, at 10 K/s using the UFS1 sensor, and at 10,000 K/s using the UFH1 sensor. (c) Selection of corresponding full heat-flow curves. All curves were normalized to the first, strongest exothermic peak and vertically offset for clarity. Dotted lines indicate the residual baseline trends after background correction.

To resolve phase evolution across the ID regime, selected-area electron diffraction (SAED) patterns were acquired from samples in the as-prepared, ID-01, ID-02, and Peak-A states, as indicated in **Figure 3**. Throughout ID-01 and ID-02, only polycrystalline fcc Ni and fcc Al are observed, with no detectable intermetallic phase formation (**Figure 5a–c, e**). The weak reflection at $|\Delta k| \approx 2.83$ 1/nm, detected in the as-prepared and ID-01 states (red circle), is attributed to preparation-induced oxide formation, as it coincides with reflections expected for $Al_2O_3$[60] and is inconsistently observed across samples. The as-prepared and ID samples further exhibit a pronounced {111} out-of-plane texture (**Figure 5a–c**), corroborating the 4D-STEM analysis. Upon reaching Peak A, several additional weak reflections emerge across the diffraction pattern (**Figure 5f**,**g**, black arrows), marking the onset of intermetallic phase formation. The reflections are best matched by a combination of $Al_9Ni_2$ and $Al_3Ni$ phases (see **Supplementary Note S3**), consistent with literature reports identifying these phases among the first products formed following interdiffusion in Ni/Al multilayers.[31,38,54–57,59,61] These observations rule out intermetallic phase formation as the origin of the calorimetric shoulder and instead point toward interdiffusion within the parent fcc phases as the dominant source of heat release.

Although no new phases form across the ID regime, the diffraction signatures evolve markedly, particularly for the Al reflections (**Figure 5f,g**). The intensity of the Al-{111} reflection changes substantially across the investigated states, becoming nearly undetectable in ID-01 before partially recovering in ID-02. At the same time, the Al-{220} and Al-{222} reflections weaken and broaden. By ID-02, their intensities decrease by approximately 20%, while their full width at half maximum (FWHM) roughly doubles (**Supplementary Figure SI 5**). In contrast, the Ni reflections show no broadening across the ID regime. Instead, the FWHM of the Ni-{220} peak decreases by approximately 30% toward ID-02, while its intensity remains largely unchanged. Furthermore, noticeable variations in peak position are observed between the investigated states. Such variations in peak positions are expected even among nominally identical samples, as the measurements were performed on separate multilayers (i.e., regions of interest), each possessing a unique microstrain and defect state. Consequently, no systematic peak shift can be identified across the ID regime.

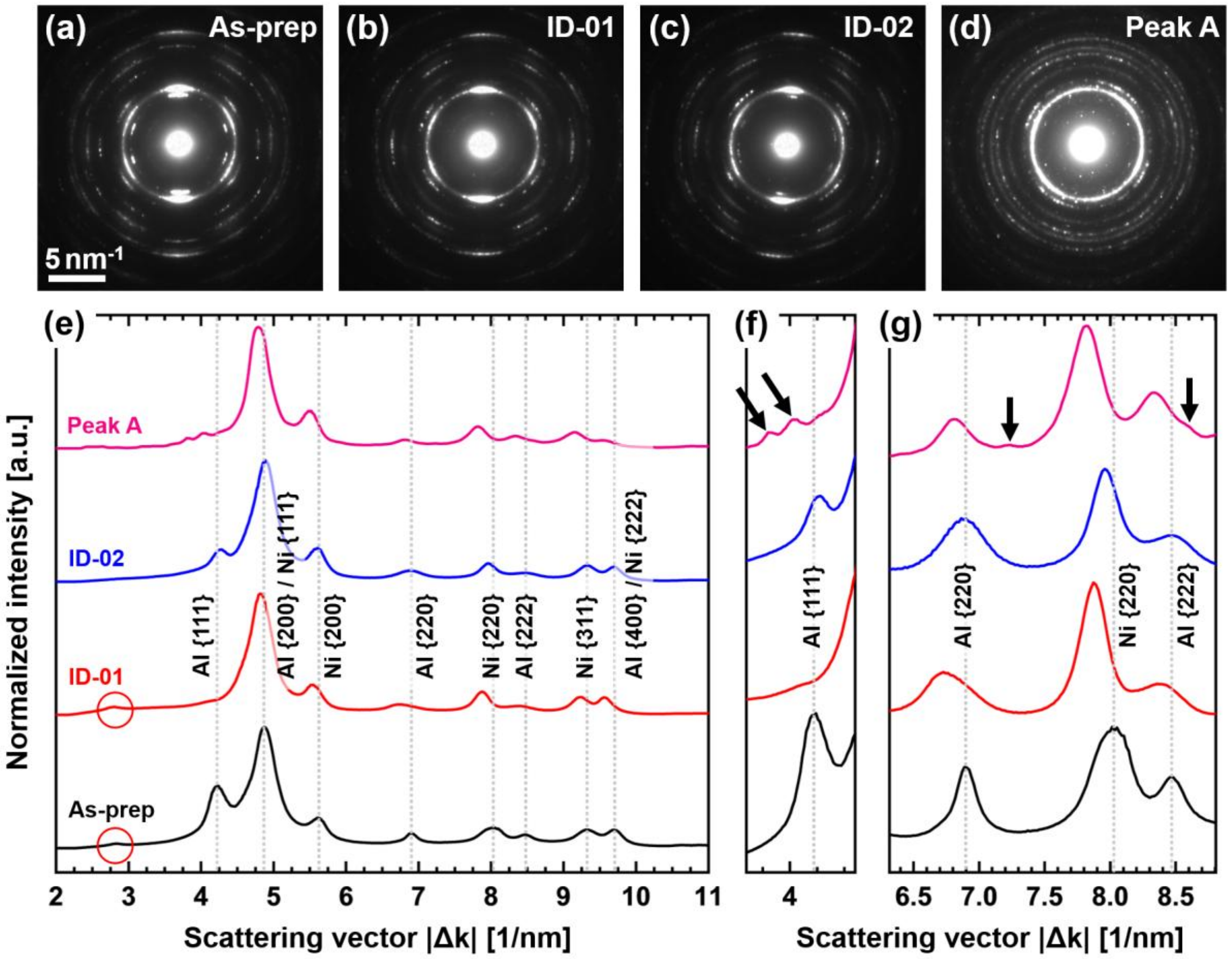

**Figure 5: SAED analysis of the as-prepared and heat-treated samples. (a–d)** SAED patterns of the as-prepared sample (a), interdiffusion stages ID-01 (b) and ID-02 (c), both obtained at 500 K/s, and the Peak-A state reached at 5000 K/s (d). **(e)** Azimuthally integrated, background-subtracted intensity profiles plotted as a function of |Δk|, normalized to the Al-{200}/Ni-{111} peak; peak assignments based on reference data.[62,63] The red circle marks a weak reflection attributed to preparation-induced $Al_2O_3$ formation. **(f,g)** Enlarged views of selected |Δk|-ranges. Black arrows indicate the first intermetallic reflections appearing in the Peak-A state.

## 2.3 Microstructural evolution across the interdiffusion regime

STEM-HAADF imaging and STEM-EELS mapping were used to track the microstructural and compositional evolution across the interdiffusion regime and into the $Al_3Ni/Al_9Ni_2$ Peak. No significant compositional changes are observed after annealing to ID-01, whereas completion of ID-02 increases the average Ni content of the Al layers from 6.3 ± 0.8 at.% in the as-prepared state to 10 ± 2 at.% (**Figure 6a**). Upon annealing to the $Al_3Ni/Al_9Ni_2$ Peak, the Ni content increases sharply to 25.7 ± 1.2 at.%, strikingly close to the stoichiometric composition of $Al_3Ni$ and consistent with its formation observed by SAED (cf. **Figure 5**). The lateral Ni distribution within the Al layers also changes markedly: whereas only limited lateral variation is present in the as-prepared state, pronounced concentration fluctuations are observed after completion of ID-02 (**Figure 6b**). We emphasize that while the chosen relative coordinate enables a facile comparison between profiles and mitigates the influence of layer roughness, it also hides layer thickness changes upon annealing and allows deviations of the curve integrals despite mass conservation.

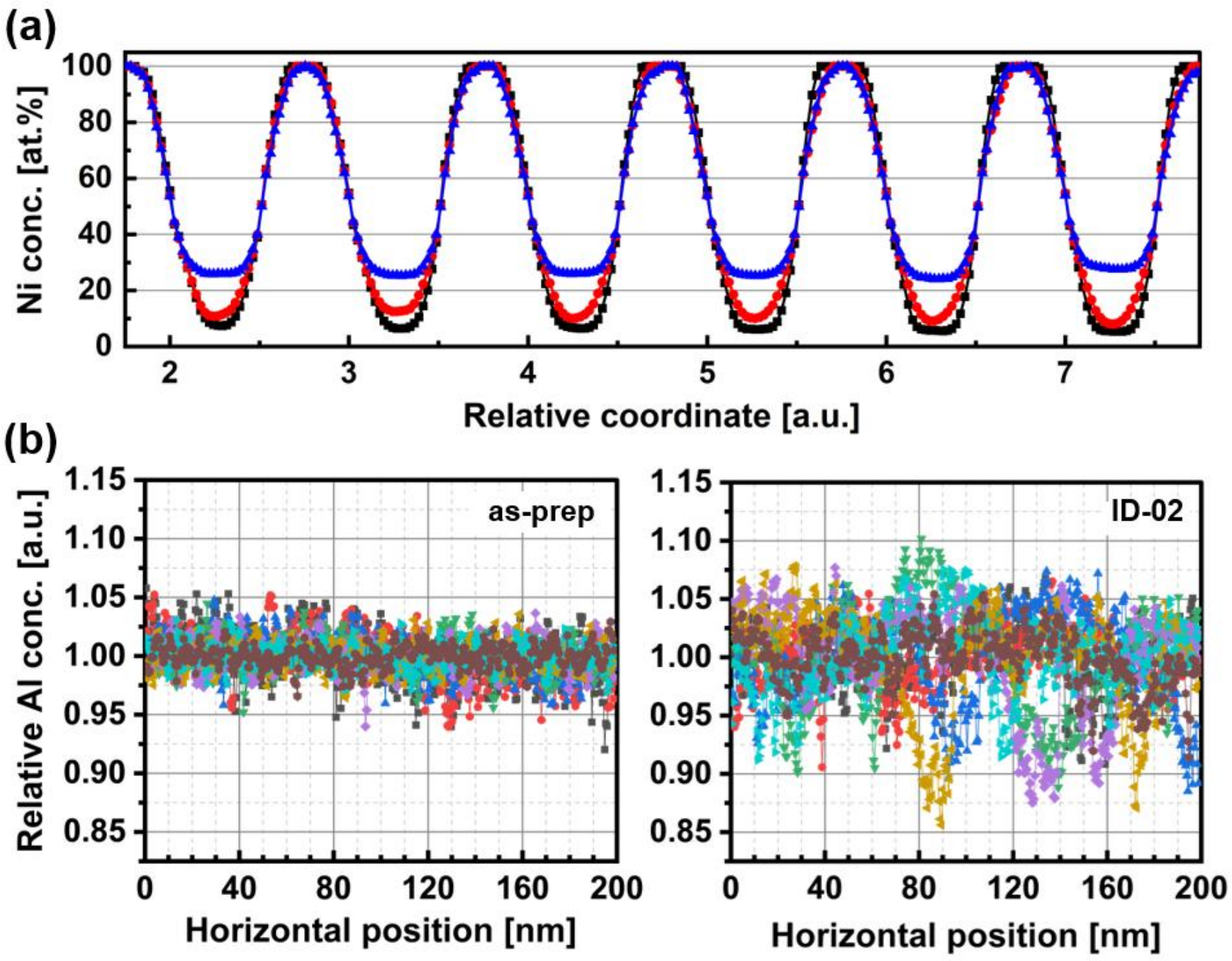

**Figure 6: Compositional evolution of the Ni/Al multilayers across the interdiffusion regime and into Peak A.** Annealed states were obtained by continuous heating at 500 K/s. (a) STEM-EELS Ni concentration profiles across the multilayer stack in the as-prepared, ID-02-completed, and Peak A states. (b) Relative Al concentration profiles along the centers of eight individual Al layers in the as-prepared and ID-02-completed states. For each profile, the Al concentration was normalized to its average value along the respective layer.

To examine the origin of these increased lateral fluctuations, selected Al layers were analyzed at higher magnification by STEM-HAADF imaging and STEM-EDX mapping (**Figure 7**). Consistent with the preceding observations, no significant changes are resolved after annealing to ID-01 (**Figure 7a**). After completion of ID-02, however, pronounced inhomogeneities emerge within the Al layers (**Figure 7b**). Bright stripes extend from the intermixed interfaces into the Al layers, often spanning the entire layer and connecting to the intermixed region at the opposite interface. Correlative EDX mapping confirms that these HAADF-bright regions correspond to local Ni enrichment (**Figure 7c**). Their characteristic spacing of ~5–10 nm closely matches the Al grain size (cf. **Figure 1**), suggesting preferential Ni enrichment along Al grain boundaries.

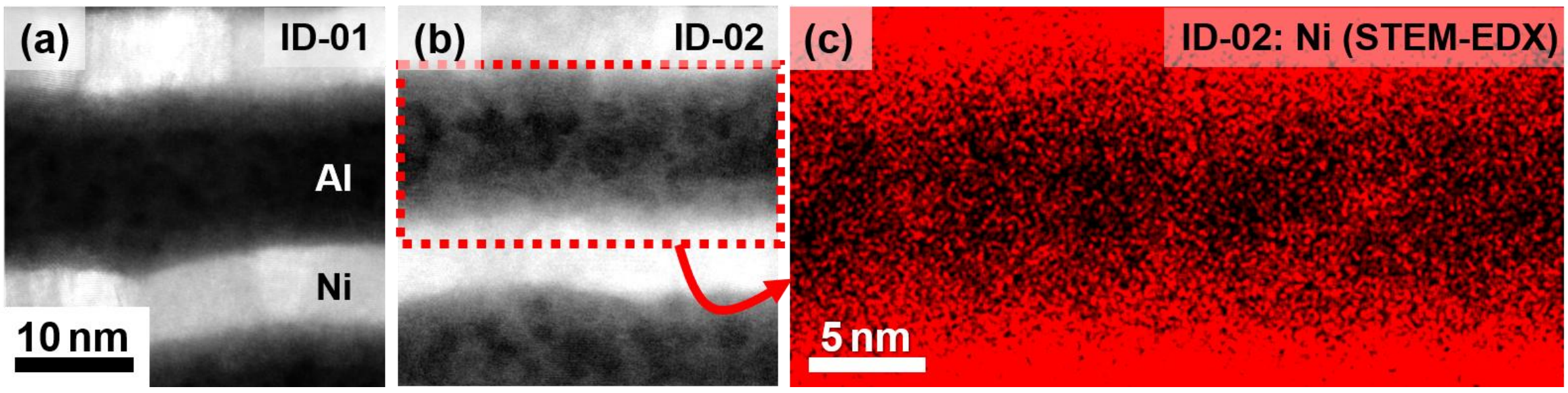

**Figure 7: Development of local Ni enrichment within Al layers across the interdiffusion regime.** (a,b) STEM-HAADF images after annealing to (a) ID-01 and (b) ID-02 at 500 K/s. (c) STEM-EDX Ni distribution map of the marked Al layer in (b).

## 2.4 Kinetic analysis of the ID regime

To quantify the kinetics of the interdiffusion regime, we applied an isoconversional Kissinger–Akahira–Sunose (KAS) analysis combining heat-flow data acquired between 0.1 and $10^4$ K/s (**Figure 8**). At very small values of the conversion parameter $\alpha$, the activation energy exhibits comparatively low values (~100 kJ/mol) and reaches a local minimum of 81 ± 24 kJ/mol near $\alpha \approx 0.02$. With increasing conversion, $E_A(\alpha)$ rises approximately linearly up to $\alpha \approx 0.2$, where it reaches a plateau with values near ~168 kJ/mol. To relate this conversion dependence to the calorimetric features, the resulting conversion scale was mapped onto a representative heat-flow curve. The local minimum coincides closely with the completion of ID-01, whereas the subsequent increase in $E_A(\alpha)$ spans ID-02 and extends into the rising flank of the $Al_3Ni/Al_9Ni_2$ peak, with the plateau coinciding with the peak maximum (**Supplementary Figure SI 6**).

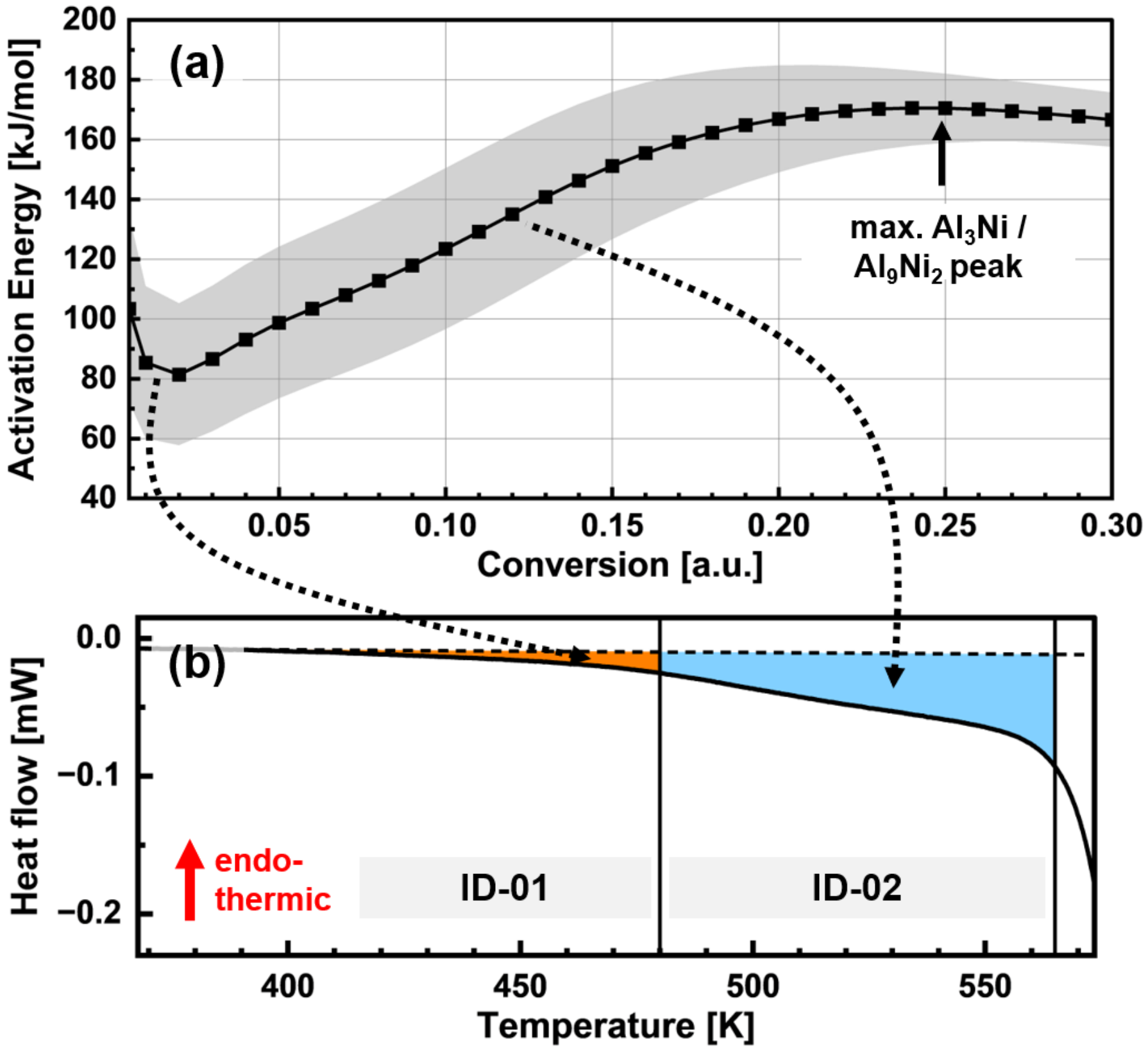

**Figure 8: KAS analysis and back-projection onto a representative heat-flow curve.** (a) Apparent activation energy $E_A(\alpha)$ derived from isoconversional KAS analysis; the shaded region indicates the standard deviation. (b) Back-projection of selected conversion levels onto a representative heat-flow curve (500 K/s). The highlighted areas correspond to the completion of ID-01 ($\alpha \approx 0.016$) and ID-02 ($\alpha \approx 0.121$). The marker at $\alpha \approx 0.248$ in (a) denotes the conversion level associated with the maximum of the $Al_3Ni/Al_9Ni_2$ peak.

# 3. Discussion

The central advance of this study is the experimental separation and quantitative characterization of the grain-boundary- and lattice-mediated transport processes governing interdiffusion in nanoscale Ni/Al multilayers. The broad exothermic shoulder preceding intermetallic formation separates into two intervals, denoted ID-01 and ID-02, most clearly at intermediate heating rates between approximately 10 and 2500 K/s (**Figure 3**). Although the distinction becomes less pronounced at lower and higher rates, the onset temperatures of both intervals shift systematically with heating rate, confirming their thermally activated character (**Figure 4**). Isoconversional KAS analysis tracks a pronounced evolution of the apparent activation energy through these intervals (**Figure 8**): the activation energy reaches a minimum at the completion of ID-01, increases continuously throughout ID-02, and approaches a plateau at conversions associated with the maximum of the $Al_3Ni/Al_9Ni_2$ peak. Such behavior is difficult to reconcile with a single dominant diffusion mechanism, for which an approximately constant activation energy with conversion would generally be expected.[64] Instead, the marked conversion dependence indicates a transition between kinetically distinct transport pathways with different activation energies. A related evolution was reported by Grapes et al. using isoconversional Friedman analysis of Al-rich Al/Ni trilayers, where the early minimum was attributed to grain-boundary interdiffusion and the subsequent increase to a growing contribution from diffusion-controlled $Al_3Ni$ formation.[44]

The activation energies resolved here allow the transport contributions within the interdiffusion regime to be distinguished more specifically. The lower value of 81 ± 24 kJ/mol lies within the range commonly associated with grain-boundary diffusion, whereas the higher value of 168 ± 17 kJ/mol approaches reported values for lattice diffusion of Ni in Al (**Table 1**). Notably, these two values also follow the empirical relation $E_{\mathrm{lattice}} \approx 2E_{\mathrm{GB}}$ for diffusion in fcc metals.[65,66] Literature values frequently fall between the two activation energies identified here, suggesting that many reported effective values reflect unresolved contributions from grain-boundary and lattice transport. Sequential activation of these pathways is not itself unexpected in a nanocrystalline material; the important result is that their contributions can be separated calorimetrically and assigned distinct kinetic barriers under heating conditions extending into the ignition-relevant regime (**Figure 3**, **Figure 8**). Together, the kinetic data support a hierarchical mechanism in which Ni transport is initially concentrated along Al grain boundaries during ID-01, before lattice diffusion from these boundaries into the grain interiors becomes increasingly important during ID-02 (**Figure 9**).

**Table 1: Activation energies reported for Ni diffusion in Al from literature.** Diffusion mechanisms are reported as hypothesized by the respective authors. *RTA*: rapid thermal annealing; *GDOES*: glow discharge optical emission spectroscopy.

| Material | Methods | Diffusion path | $E_A$ [kJ/mol] | Ref. |
|---|---|---|---|---|
| Ball-milled Ni–Al, Al grains 10–20 nm | In-situ TEM + Sauer–Freise | GB + lattice | 73–111 | [67] |
| Ni/Al multilayers, Λ = 20–200 nm | hot-plate experiments + ignition-threshold modeling | GB | 77.3 ± 1.3 | [68] |
| Ni/Al multilayers, Λ = 30 nm, Al:Ni = 1:1 | In-situ X-ray reflectivity | GB | 80 ± 19 | [69] |
| **Ni/Al multilayers, Λ = 20 nm, Al:Ni = 1:1** | **FDSC + KAS** | **GB** | **81 ± 24** | **This work** |
| Ni/Al multilayers, Λ = 25 nm, Al:Ni = 1:1 | In-situ X-ray reflectivity | GB | 92 ± 7 | [69] |
| Al(41 nm)/Ni(18 nm)/Al(41 nm), Al:Ni = 3:1 | Nanocalorimetry (Friedman), in situ TEM | GB | 113 ± 4 | [44] |
| Ni/Al bilayer, Al/Ni = 1µm/1µm | RTA + GDOES (Boltzmann–Matano) | GB | 127 ± 5 | [70] |
| Ni-coated Al wires, mm-sized Al grains | Resistometric impurity diffusion | lattice | 146 ± 6 | [71] |
| **Ni/Al multilayers, Λ = 20 nm, Al:Ni = 1:1** | **FDSC + KAS** | **lattice** | **168 ± 17** | **This work** |

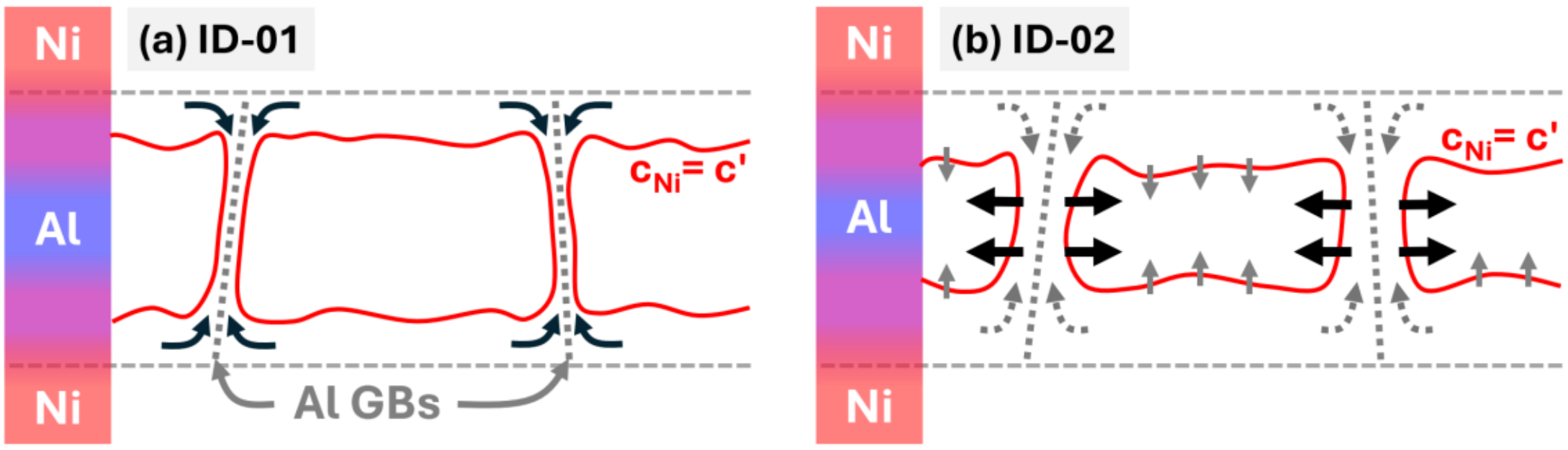

**Figure 9: Schematic of the proposed hierarchical interdiffusion mechanism.** (a) ID-01, where Ni transport is predominantly confined to Al grain boundaries (grey dashed lines). The red contour denotes a constant Ni concentration ($c_{Ni} = c_{Ni}'$). (b) ID-02, where grain-boundary diffusion is supplemented by lattice diffusion into the Al grain interiors, leading to redistribution of Ni throughout the grains. Black arrows indicate the dominant diffusion pathways at each stage, while grey solid arrows represent comparatively minor bulk diffusion.

Electron microscopy at defined stages across the interdiffusion regime provides an independent microstructural test of this kinetic interpretation. SAED analysis shows that the broad calorimetric shoulder is not associated with detectable intermetallic-phase formation, indicating that the heat released during ID-01 and ID-02 arises predominantly from interdiffusion within the parent fcc phases (**Figure 5**). STEM-EELS mapping further resolves the associated compositional evolution: no significant changes are observed after ID-01, whereas completion of ID-02 increases the average Ni content of the Al layers from $6.3 \pm 0.8$ to $10 \pm 2$ at.% and produces pronounced lateral concentration fluctuations (**Figure 6**). Higher-resolution HAADF imaging and EDX mapping reveal corresponding Ni-enriched stripes extending from the intermixed interfaces into the Al layers, frequently across their entire thickness (**Figure 7**). Although enrichment of individual grain boundaries cannot be assigned unambiguously within the spatial-resolution and projection limitations of the STEM-EDX measurements, the characteristic spacing of these features, approximately 5–10 nm, closely matches the Al grain size. This correspondence supports a scenario in which grain boundaries first provide rapid pathways across the Al layers and subsequently act as distributed sources for lattice diffusion into the adjacent grain interiors. The observations are also consistent with atom-probe studies identifying Al grain boundaries as dominant transport pathways in Ni/Al diffusion couples.[17,18] Together, the kinetic and microscopy results support the Al grain-boundary network as the dominant low-barrier pathway governing the onset and spatial progression of interdiffusion.

This mechanistic picture has direct implications for tailoring heat release in reactive multilayers. Increasing the Al grain size, and thereby reducing the density of through-thickness grain boundaries, should diminish the contribution of ID-01 and may delay or broaden the subsequent heat release. This interpretation is consistent with molecular-dynamics calculations showing that increasing structural disorder and grain-boundary density increases the reaction rate and reaction-front velocity in Al/Ni multilayers.[21,22] A more targeted, although presently less established, strategy would be the selective decoration of Al grain boundaries with minor alloying elements that reduce transport along these pathways. Atomistic calculations for Al–Mg demonstrate that solute segregation can substantially reduce grain-boundary mass transport in Al by increasing the corresponding diffusion barrier, providing proof of principle for such solute-mediated grain-boundary blocking.[72] In reactive Ni/Al multilayers, Nb alloying has already been shown to form an in-situ diffusion barrier and reduce the heat-flow rate by more than 50%, while solid-solution alloying of the Ni layers with Cu or Pt can change the rate-limiting intermixing mechanism and strongly modify propagation velocity and temperature.[73,74] These studies do not yet demonstrate selective passivation of the Al grain-boundary network, but they establish alloying as a viable route for manipulating transport and reaction kinetics. The present workflow now provides a means of testing such interventions mechanistically: changes in the low-conversion activation energy, the heat partition between ID-01 and ID-02, and the corresponding arrested microstructures can reveal whether a modification selectively suppresses the fast grain-boundary pathway, also affects lattice transport, or more broadly alters the phase sequence or reaction enthalpy.

Beyond the specific Ni/Al system, the combination of flash calorimetry, isoconversional kinetic analysis, and state-resolved electron microscopy establishes a transferable framework for disentangling rapid transformations that overlap under conventional measurement conditions. By separating these transformations into kinetically distinct regimes, quantifying their associated barriers, and testing the resulting assignments in arrested transient microstructures, the approach directly connects calorimetric signatures to their underlying mechanisms. This provides a basis for systematically determining how microstructural design modifies individual transport and reaction pathways in reactive multilayers and, more broadly, in materials driven far from equilibrium, where diffusion, clustering, ordering, and phase formation processes are closely coupled in time.

# 4. Methods

### Ni/Al multilayer deposition

Ni/Al multilayers were deposited by DC magnetron sputtering onto Si (100) substrates using a von Ardenne CS400 system. High-purity targets (Ni: 99.98 wt.%, Al: 99.999 wt.%) were used (FHR Anlagenbau GmbH). Layer thicknesses were controlled via calibrated deposition rates, yielding a bilayer periodicity of $\Lambda = 20$ nm (≈ 8 nm Ni / 12 nm Al) and an overall composition of $Ni_{50}Al_{50}$ (at.%). Multilayer thin films with total thicknesses of 5 µm and 1 µm were prepared to accommodate different calorimetric conditions. Free-standing films were obtained by mechanically delaminating the multilayers from the substrate. Deposition parameters and setup details are provided in **Supplementary Table SI 1**.

### Differential scanning calorimetry

Conventional DSC measurements were performed on stacked free-standing films (t = 5 µm, 1–4 mg) in Al pans using a Mettler-Toledo DSC 3 equipped with a Huber TC45 intracooler. Measurements were conducted between 253–853 K at heating rates of 0.1–1 K/s under a continuous flow of high-purity Ar (20 mL/min). A second heating scan of the reacted sample served as baseline. Temperature and enthalpy were calibrated using standard metals for each heating rate and crucible configuration.

### Fast differential scanning calorimetry

FDSC measurements were conducted using a Mettler-Toledo FDSC 2+ equipped with a Huber TC100 intracooler under Ar atmosphere (20 mL/min). Samples (≈ 10–100 ng, estimated following Ivanisenko et al.[75]) were cut from free-standing films and placed onto chip sensors (UFS1 or UFH1). Heating rates spanning six orders of magnitude (0.1–$10^4$ K/s) were applied using logarithmic increments ($1\times10^n$, $2.5\times10^n$, $5\times10^n$, $10\times10^n$ K/s, n = 0–3). Lower rates (≤ 500 K/s) were measured on UFS1 sensors, while higher rates were achieved using UFH1 sensors. Multilayers with a thickness of 5 µm were used for heating rates up to 1000 K/s, while thinner films (t = 1 µm) were used at higher rates to avoid self-propagating reactions.[38] Samples were heated from 223 K to maximum temperatures between 773 K and 1073 K depending on sensor type and sample thickness. Each measurement was repeated at least three times, and baselines were obtained from subsequent scans of the reacted samples. Temperature calibration was performed for each chip sensor using standard metals following Monnier et al.,[76] and thermal lag was corrected following Riegler et al.[38]

### Kissinger–Akahira–Sunose analysis

Kinetics of the interdiffusion regime were evaluated using an isoconversional Kissinger–Akahira–Sunose (KAS) analysis.[64,77] This integral approach enables the extraction of conversion-dependent activation energies without assuming a specific reaction model or requiring the reaction rate to reach a maximum, making it particularly suitable for the broad exothermic interdiffusion shoulder. The conversion parameter α was obtained by integrating the heat-flow signal from the onset of the exothermic response, using an incremental step size of $\Delta\alpha = 0.01$. For each heating rate β, the temperature corresponding to a given conversion level ($T_\alpha$) was determined and used to construct KAS plots of $\ln(\beta/T_\alpha^2)$ versus $1/T_\alpha$. Apparent activation energies were obtained from the slope of linear fits at constant α. Further details are provided in **Supplementary Note S4**.

### FDSC-based annealing and TEM sample preparation

Defined reaction states were accessed by FDSC annealing followed by rapid quenching ($10^4$ K/s) using a UFH1 sensor. Samples were heated at 500 K/s to probe the interdiffusion regime or at 5000 K/s to access early intermetallic formation. The lower heating rate enables clear separation of the ID-01 and ID-02 regimes, while the higher rate maximizes contrast to low-rate reference experiments (0.33 K/s). Corresponding heat-flow traces confirm reproducible arrest in the targeted states (**Supplementary Figure SI 7**). After annealing, films were removed from the sensor and transferred to SEM stubs for focused ion beam (FIB) preparation. TEM lamellae were prepared using a FEI Helios NanoLab 600 following standard procedures.[78,79] Final polishing was performed at 5 keV to minimize ion damage, followed by low-energy ion milling using a PIPS system (4D-STEM; 300eV acceleration voltage, 28µA beam current, +/− 10° tilt, 40s total) or a NanoMill system (STEM-HAADF, TEM-SAED; 900eV acceleration voltage, ~70µA beam current, +/− 10° tilt, 15–20 min total).

### 4DSTEM and STEM-EELS

4DSTEM was performed on an image-corrected FEI Titan 80–300 ETEM operated at 300 kV using a convergence semi-angle of 1 mrad and a beam current of 30 pA. Data were acquired using a custom acquisition script[80] with a Gatan UltraScan 1000XP camera binned by 8 and a real-space pixel size of 0.5 nm. A camera length of 48 mm was used, resulting in a reciprocal-space sampling of 0.136 $nm^{-1}px^{-1}$. STEM-EELS was performed on the same microscope using a Gatan Quantum 965 ER imaging filter. Spectra were acquired with a convergence semi-angle of 10 mrad, a beam current of 100 pA, and a camera length of 38 mm, corresponding to a collection semi-angle of 47 mrad. The energy dispersion was set to 1 eV per channel. Atomic concentrations were determined using the elemental-quantification routine in Gatan DigitalMicrograph 3.62, including power-law background subtraction and deconvolution based on the corresponding low-loss spectrum. For layer-referenced comparisons, the resulting concentration maps were transformed to an interface-aligned relative coordinate by mapping successive 50 at.% Al crossings, taken as the local interface positions, to fixed coordinates and linearly interpolating between them, as detailed in **Supplementary Note S1**.

### TEM, STEM-HAADF and EDX analysis

TEM, SAED, and HAADF-STEM imaging were performed on an aberration-corrected JEOL ARM200F operated at 200 kV. HAADF-STEM imaging was carried out using a nominal convergence semi-angle of 25 mrad. Additional HAADF-STEM imaging and EDX analysis were performed using a probe-corrected Titan G2 80–200 microscope operated at 200 kV and equipped with a Super-X detector. Imaging on the Titan was carried out using a convergence semi-angle of 24.7 mrad and a collection angle range of 69–200 mrad.

# 5. Supplementary Materials

## 5.1 Supplementary Note S1: Interface-aligned coordinate transformation

Local variations in interface position and individual layer thickness complicate direct comparison and averaging of the STEM-EELS concentration maps, because pixels at the same absolute vertical coordinate do not represent equivalent positions within a layer. To minimize this geometric contribution, the maps were transformed from an absolute spatial coordinate to an interface-aligned relative coordinate (**Figure SI 1**). For each pixel column, i.e., fixed lateral position, the vertical points $x_i$ at which the Al concentration crosses 50 at.% were identified. These points were mapped to integer values of the relative coordinate for Al-to-Ni interfaces and half-integer values for Ni-to-Al interfaces hence the relative coordinate unit corresponds to one bilayer. The relative coordinate $u$ between two adjacent crossings $x_i$ and $x_{i+1}$ was obtained by linear interpolation:

$$u(x) = u_i + \frac{x - x_i}{x_{i+1} - x_i}(u_{i+1} - u_i).$$

This transformation locally stretches or compresses the coordinate between adjacent interfaces, thereby mapping the individual layers to a common relative thickness and aligning corresponding interfaces across the concentration map.

The resulting roughness-normalized maps enable standardized evaluation of average concentrations within equivalent layer-center regions and of lateral concentration variations along the Al layer centers. Applying the same procedure to the different arrested reaction states further facilitates comparison between independently prepared TEM specimens while reducing the influence of site-specific interface roughness and variations in individual layer thickness. The transformation intentionally removes information on absolute local layer dimensions, interface morphology, and annealing-induced thickness changes. Because the coordinate is locally rescaled, curve integrals in the transformed coordinate are not required to remain constant despite mass conservation in real space. The method was therefore used only for layer-referenced concentration comparisons and not for determining absolute interface widths, diffusion distances or layer thickness changes.

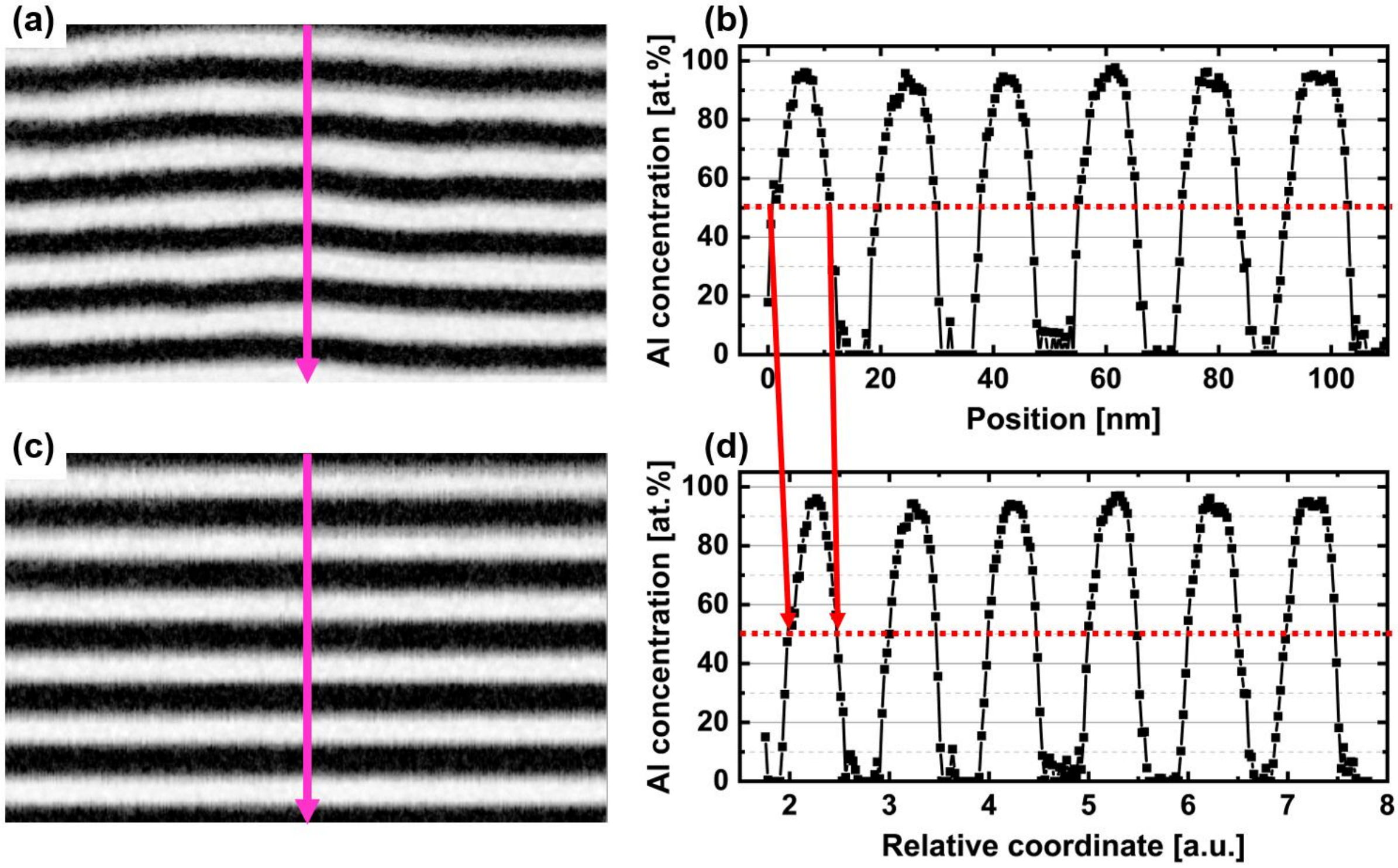

**Figure SI 1: Interface-aligned transformation of an Al concentration map.** (a) Al concentration map before roughness normalization. (b) Concentration profile along the pink arrow in (a). Successive 50 at.% Al crossings are assigned alternating integer and half-integer relative coordinates for Al-to-Ni and Ni-to-Al interfaces, respectively, as illustrated by the red arrows. Coordinates between adjacent crossings are linearly interpolated. (c) Resulting roughness-normalized concentration map. (d) Concentration profile along the pink arrow in (c).

## 5.2 Supplementary Figure S2

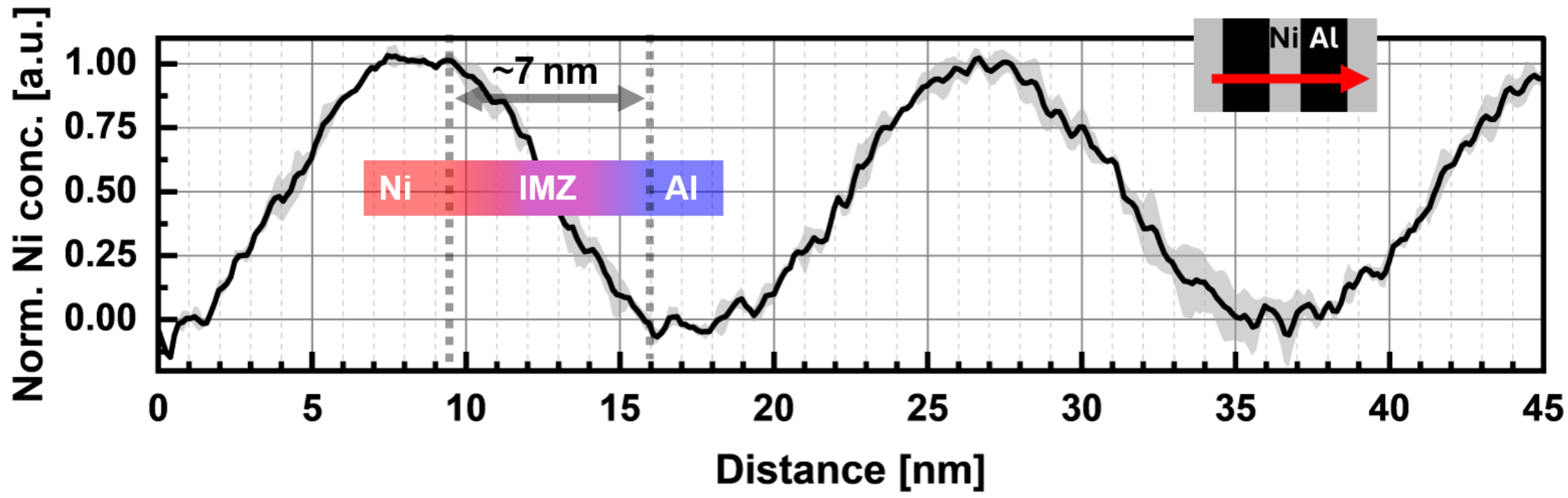

**Figure SI 2: STEM-EDX line profile across the as-prepared Ni/Al multilayer.** Profiles were extracted from well-defined regions with locally parallel layer interfaces and aligned to the local interface positions before averaging. The shaded band indicates the resulting spread across the analyzed regions of interest. The averaged Ni signal was normalized to representative plateau values in the centers of the Al and Ni layers, assigned normalized values of 0 and 1, respectively. One representative apparent intermixing zone (IMZ), defined as the region between the end of one signal plateau and the onset of the adjacent one, is marked by grey dashed lines.

## 5.3 Supplementary Note S2: Premixing in sputter-deposited Ni/Al multilayers

The as-deposited multilayers exhibit two characteristic features of premixing: (i) apparent intermixed interface regions (IMZs) at the Ni/Al interfaces and (ii) pronounced Ni enrichment throughout the Al layers, whereas the Ni layers remain comparatively free of Al (**Figure 2**). Similar premixed states have been widely reported in sputter-deposited Ni/Al multilayers[46–52] and become particularly pronounced for small bilayer periodicities.[48,51]

At first sight, the observed asymmetry appears consistent with thermally activated interdiffusion, as the diffusivity of Ni in Al exceeds that of Al in Ni by several orders of magnitude.[61,81,82] Furthermore, the negative enthalpy of mixing of the Ni–Al system thermodynamically favors the formation of metastable intermixed regions.[46,83] However, several observations indicate that thermally activated diffusion alone may not fully account for the premixed state. Most notably, the concentration gradients across the interfaces appear comparatively symmetric (**Figure SI 3**), whereas diffusion-controlled intermixing would be expected to generate markedly skewed concentration profiles due to the strong diffusivity asymmetry.[50,52,61,84,85] In addition, the Ni distribution within the Al layers does not exhibit an obvious correlation with the underlying grain structure (**Figure 6b**), despite grain boundaries being expected to provide preferred diffusion pathways during low-temperature interdiffusion.[17,18]

Ballistic mixing during sputter deposition therefore likely contributes to the formation of the premixed state. Atomistic simulations and experimental studies of sputter-deposited metallic multilayers have shown that energetic deposited atoms, reflected neutrals, and recoil events generated during film growth can induce interlayer mixing even in the absence of significant thermal diffusion.[86–89] Such mechanisms are expected to produce asymmetric mixing when the two constituents differ strongly in atomic mass. Consistent with this expectation, anomalously large and highly asymmetric intermixing has been reported for a variety of sputter-deposited Al/transition-metal systems, including Al/Ni, where transition-metal penetration into Al substantially exceeds Al penetration into the transition-metal layer.[53]

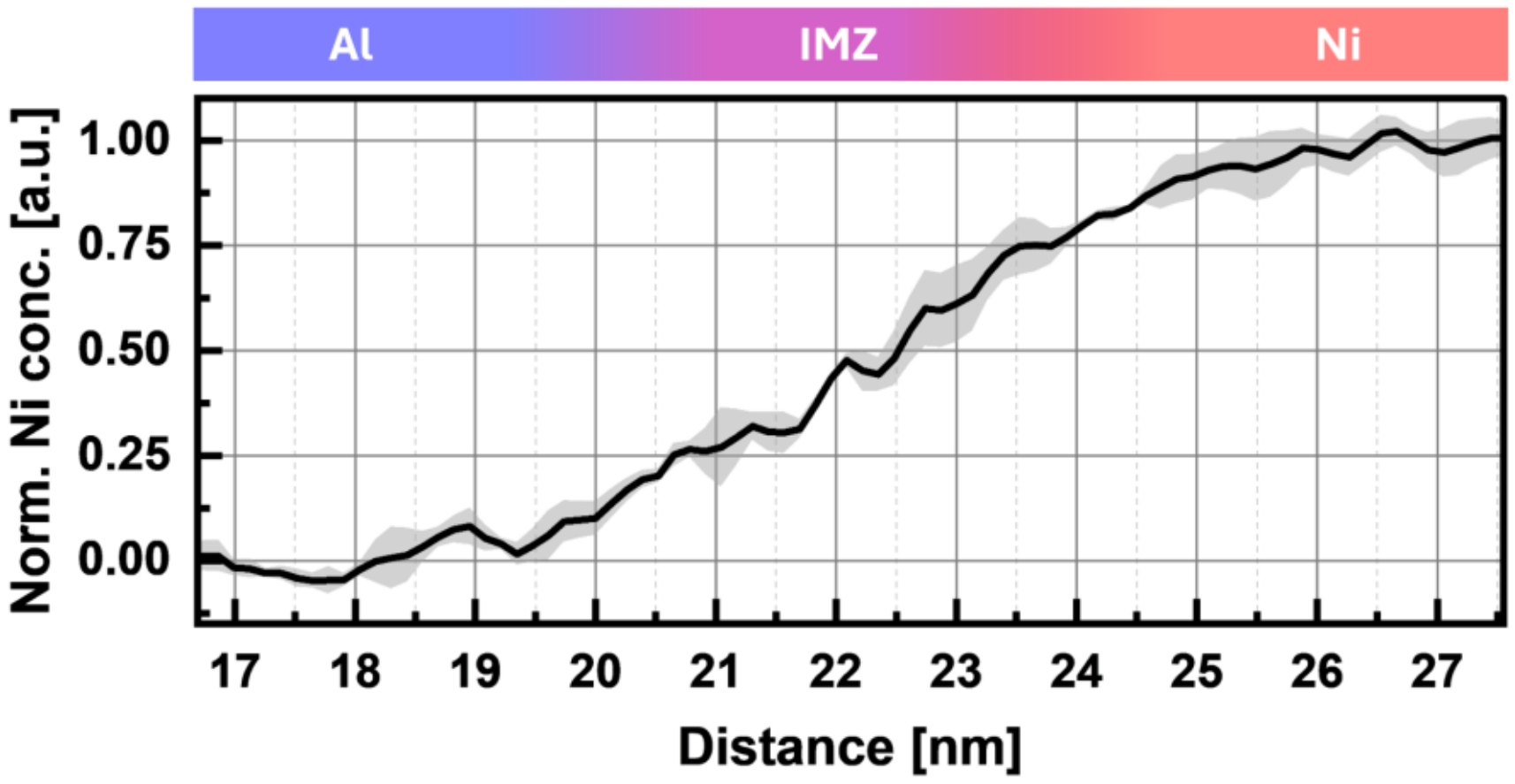

**Figure SI 3: Averaged, normalized STEM-EDX Ni profile across Al-to-Ni interfaces in the as-prepared state**. The shaded band indicates the spread across the aligned regions of interest.

Taken together, the available evidence suggests that the premixed state observed here most likely results from a combination of thermodynamic driving forces for Ni–Al mixing and ballistic intermixing processes operating during sputter deposition. The present data do not allow these contributions to be quantitatively separated.

## 5.4 Supplementary Note S3: Identification of $Al_9Ni_2$ and $Al_3Ni$ reflections in the Peak-A diffraction pattern

Phase identification at Peak A was performed by comparing the low-scattering-vector region of the azimuthally integrated SAED intensity profile with calculated diffraction peak positions and relative intensities for the stable $Al_3Ni$ and metastable $Al_9Ni_2$ phase (**Supplementary Figure SI 4**). For $Al_3Ni$, diffraction data were calculated from the reported crystal structure.[90,91] For $Al_9Ni_2$, no suitable crystallographic file was available. Therefore, following the approach of Blobaum et al.,[59] a structural model was derived from $Al_9Co_2$ (JCPDS 30-0007), which shares the same space group ($P2_1/a$) and similar lattice parameters. In accordance with the reported $Al_9Ni_2$ structure, Al atoms occupying the Wyckoff a sites in $Al_9Co_2$ were removed prior to the diffraction calculation.

The calculated line intensities correspond to relative powder diffraction intensities derived from the respective crystal structures. Direct comparison with the experimental SAED intensities should therefore be made with caution, as the intermetallic phases nucleate within a highly textured and strained multilayer microstructure. Preferred crystallographic orientations, finite grain statistics, and local strain variations may substantially modify the relative peak intensities observed experimentally. Consequently, phase identification is based primarily on the agreement of peak positions and the overall reflection pattern rather than on exact intensity matching.

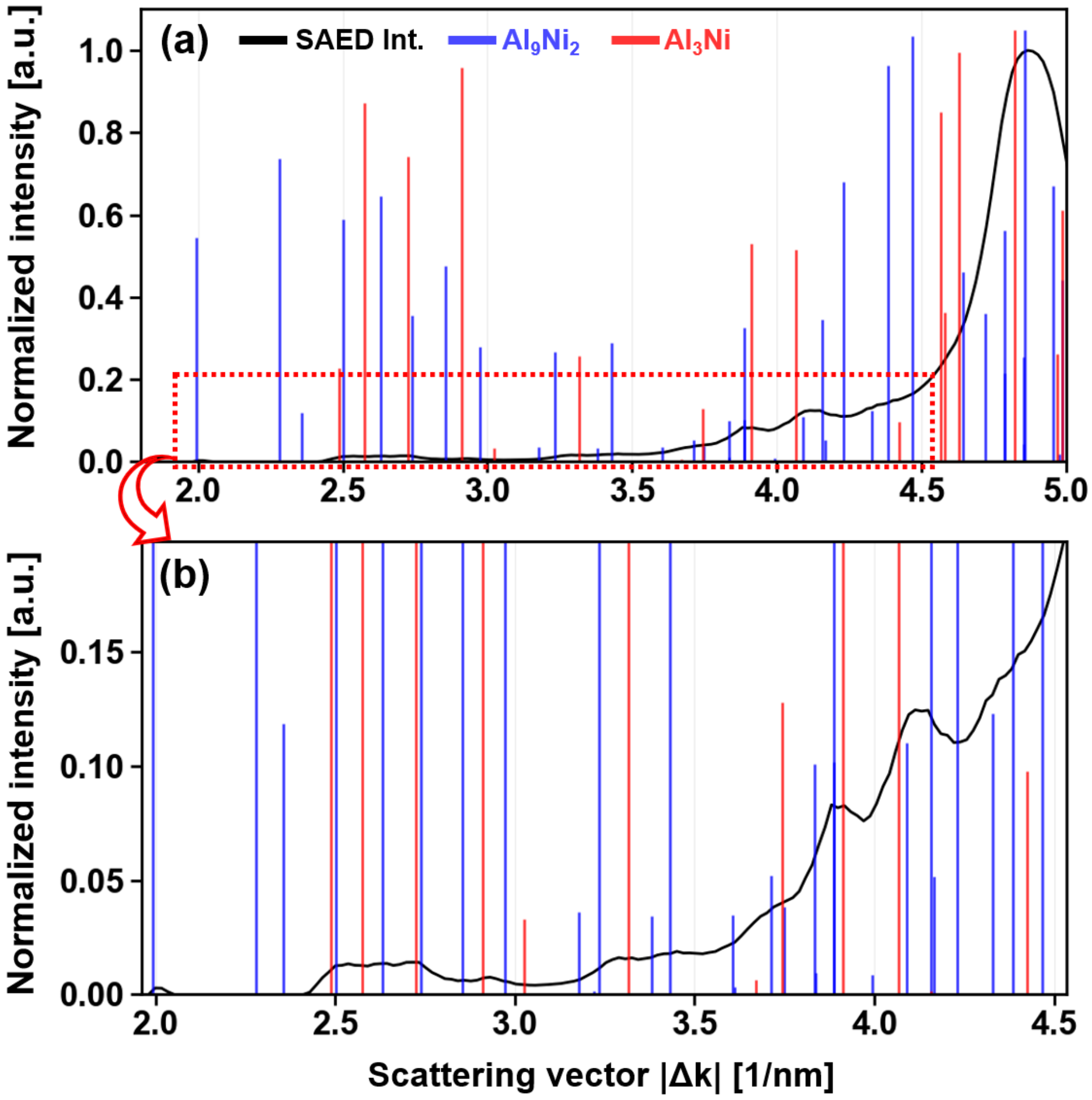

**Figure SI 4: Comparison of experimental SAED data with calculated diffraction patterns for $Al_3Ni$ and $Al_9Ni_2$. (a)** Azimuthally integrated intensity profile of the Peak-A state in the low-scattering-vector range. Vertical lines indicate expected reflection positions, while line heights represent relative powder diffraction intensities calculated from the corresponding crystal structures. The observed reflections are best matched by a combination of $Al_9Ni_2$ and $Al_3Ni$ phases. **(b)** Enlarged view of the boxed region in (a).

## 5.5 Supplementary Figure S3: Evolution of peak intensities and FWHM of selected peaks

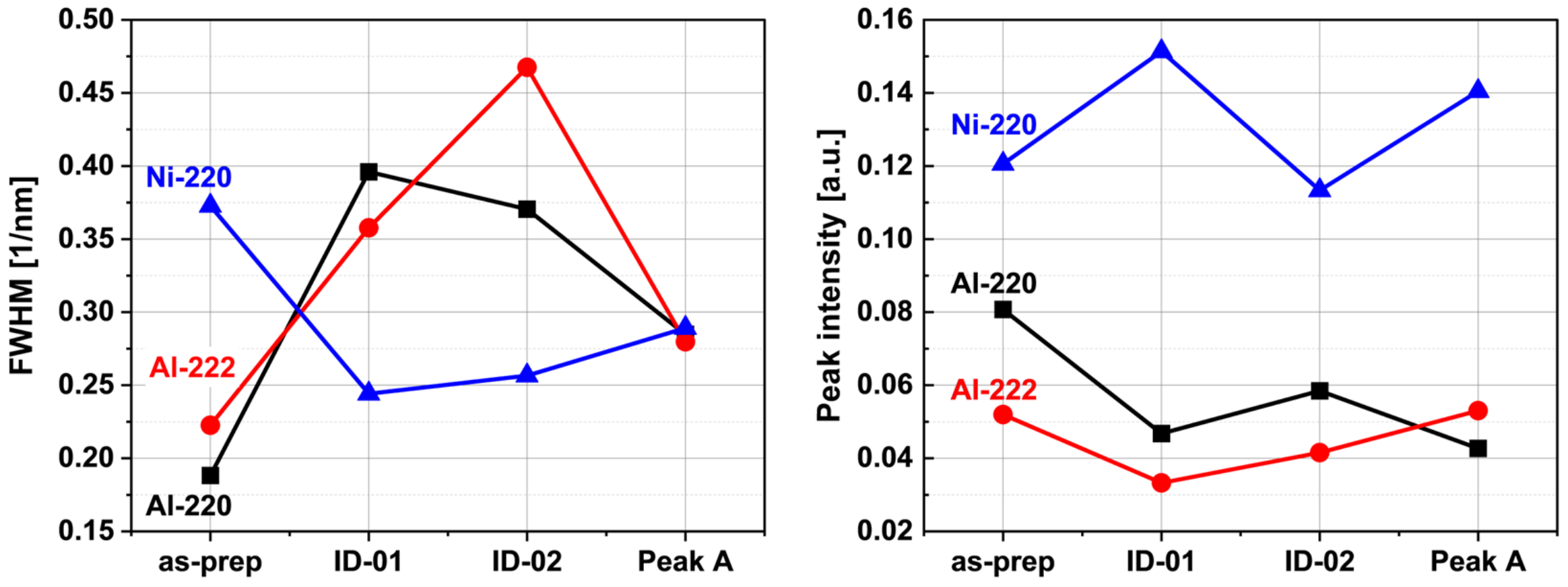

**Figure SI 5: Evolution of (a) peak intensities and (b) full widths at half maximum (FWHM) for the Ni-{220}, Al-{220}, and Al-{222} reflections** in the as-prepared state, during interdiffusion stages ID-01 and ID-02, and in the Peak-A state, extracted from the azimuthally integrated profiles shown in **Figure 5e,g**.

## 5.6 Supplementary Figure S4: Determination of the conversion corresponding to the $Al_3Ni/Al_9Ni_2$ peak maximum

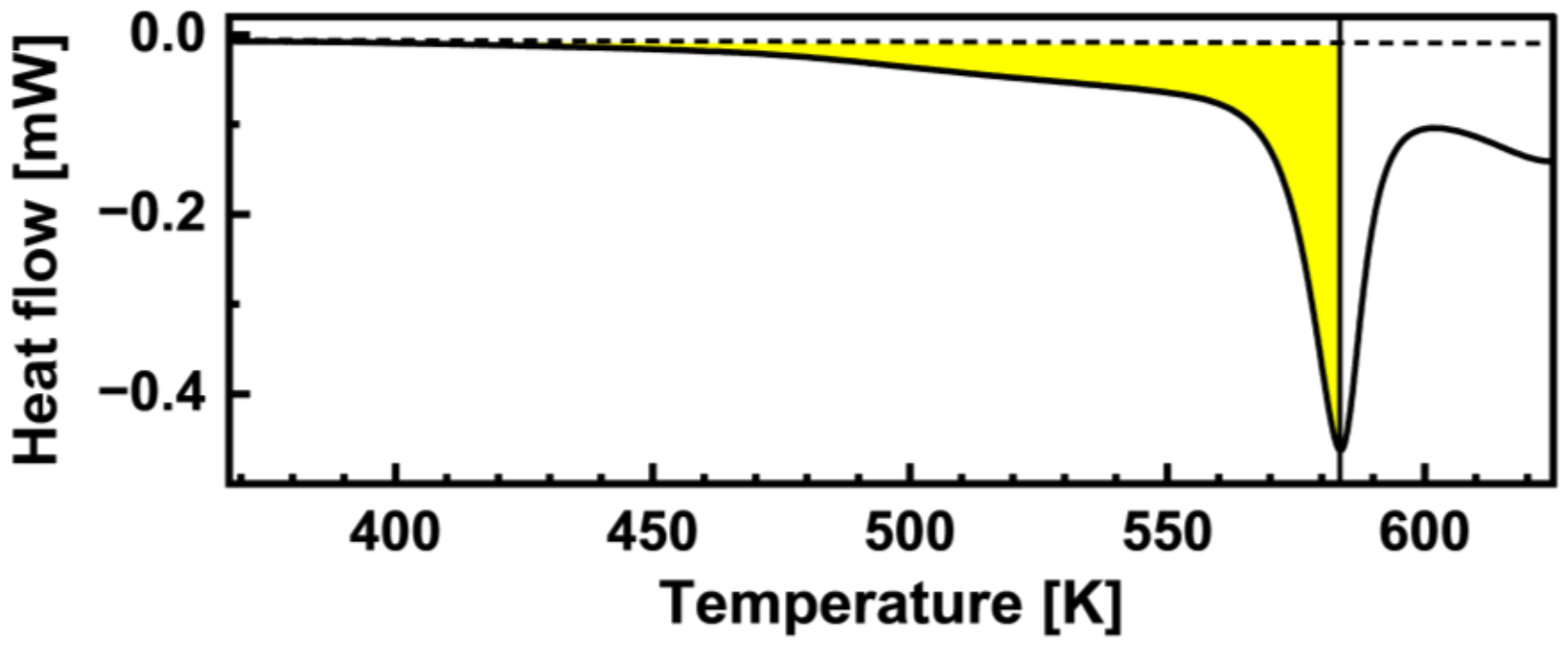

**Figure SI 6: Back-projection of the conversion level associated with the maximum of the $Al_3Ni/Al_9Ni_2$ peak.** The shaded area indicates the cumulative heat-flow integral up to the peak maximum in a representative heat-flow curve measured at 500 K/s. Normalization by the total integrated heat release yields $\alpha \approx 0.248$, corresponding to the marker shown in the KAS analysis in **Figure 8**.

## 5.7 Supplementary Figure S5: FDSC traces confirming arrest at defined reaction states

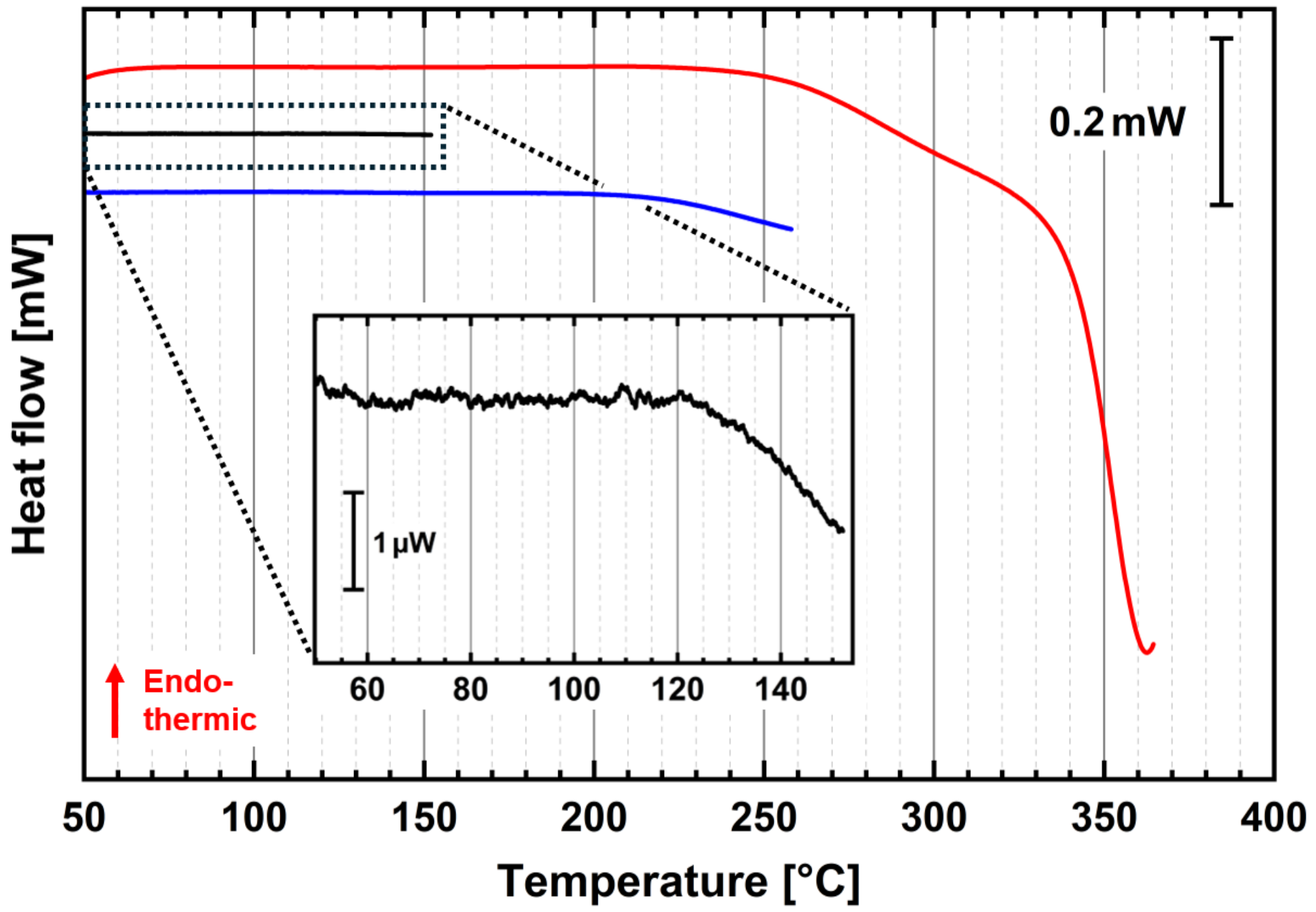

**Figure SI 7: Verification of state arrest for ex-situ microscopy.** Representative heat-flow curves of equimolar Ni/Al multilayers ($\Lambda$ = 20 nm) heated to the target states corresponding to ID-01 (black), ID-02 (blue), and the $Al_3Ni/Al_9Ni_2$ peak (red). ID-01 and ID-02 were approached at 500 K/s, whereas Peak A was reached at 5000 K/s. At the respective arrest points, heating was terminated and the samples were quenched at 10,000 K/s to preserve the corresponding transient microstructures for ex-situ characterization.

## 5.8 Supplementary Table S1: Deposition parameters

**Table SI 1: Deposition parameters for the fabrication of Ni/Al multilayer films with Λ = 20 nm**

| target material | Ar flow [mL min$^{-1}$] | substrate temperature [K] | power [W] | working pressure [$10^{-3}$ mbar] | rate [nm/s] | ind. layer thickness [nm] | total thickness [µm] |
|---|---|---|---|---|---|---|---|
| Ni | 30 | 298 | 200 | 5 | 0.33 | 8 | 5 |
| Al | | | | | 0.43 | 12 | |
| Ni | 30 | 298 | 200 | 5 | 0.33 | 8 | 1 |
| Al | | | | | 0.43 | 12 | |

## 5.9 Supplementary Note S4: Isoconversional KAS Analysis of the Interdiffusion Regime

Peak-based analyses such as the Kissinger method[64,92] require a well-defined rate maximum and are therefore not applicable to the low-temperature interdiffusion (ID) regime in the present Ni/Al multilayers, which appears as a broad shoulder in the heat-flow signal. To access the kinetics of this stage, an isoconversional Kissinger–Akahira–Sunose (KAS) approach[64,77] was employed, allowing the apparent activation energy to be evaluated as a function of reaction progress, expressed in terms of a dimensionless conversion parameter α, without assuming a specific reaction model.

The KAS method is based on the general rate equation

$$\frac{d\alpha}{dt} = k(T)\, f(\alpha)\,, \qquad \text{with} \qquad k(T) = A \exp\left(-\frac{E_A}{RT}\right),$$

which, for a constant heating rate $\beta = dT/dt$, leads to the linear relationship for a fixed conversion $\alpha$:

$$\ln\left(\frac{\beta}{T_\alpha^2}\right) = -\frac{E_A(\alpha)}{R\, T_\alpha} + \text{const.}$$

Here, $T_\alpha$ denotes the temperature at which a given conversion level is reached.

The conversion parameter $\alpha(T)$ was obtained by numerical integration of the baseline-corrected heat-flow signal:

$$\alpha(T) = \frac{\int_{T_0}^{T} \dot{q}\,(T')\,dT'}{\int_{T_0}^{T_f} \dot{q}\,(T')\,dT'} \;,$$

where $\dot{q}(T)$ is the measured heat-flow rate, $T_0$ the onset temperature, and $T_f$ the temperature at which the reaction is complete. Identical integration limits were used for all heating rates. Conversion-dependent activation energies $E_A(\alpha)$ were determined from linear fits of $\ln(\beta/T_\alpha^2)$ versus $1/T_\alpha$ at fixed $\alpha$, using an incremental step size of $\Delta\alpha = 0.01$.

# 6. Data availability

All data generated or analyzed during this study are included in this published article and its Supplementary Information files.

# 8. Acknowledgements

The authors thank Sebastian Matthes and Konrad Jaekel for their support with multilayer preparation at TU Ilmenau. Support from the Center of Micro- and Nanotechnologies (ZMN), a DFG-funded core facility at TU Ilmenau, is gratefully acknowledged. The authors further acknowledge the use of equipment at the Collaborative Laboratory and User Facility for Electron Microscopy (CLUE, www.clue.physik.uni-goettingen.de) and the FIB instrument of the Correlative Microscopy and Tomography (CoMiTo) core facility at Saarland University. This work further contains results obtained from experiments performed at the Ernst Ruska-Centre (ER-C) for Microscopy and Spectroscopy with Electrons at the Forschungszentrum Jülich (FZJ) in Germany (http://dx.doi.org/10.17815/jlsrf-2-68). The ER-C beam-time access was provided via the DFG Core Facility Project FZJ_IEK-2_PN1.

# 9. Author contributions

**S.S.R.:** Conceptualization, Data curation, Formal analysis, Investigation (lead), Methodology, Validation, Visualization, Writing – original draft, Writing – review & editing. **I.G.:** Conceptualization, Funding acquisition, Methodology, Project administration, Supervision (lead), Writing – review & editing. **N.J.P.:** Investigation, Writing – review & editing. **A.T.:** Formal analysis, Writing – review & editing. **T.M.:** Investigation, Visualization, Writing – review & editing. **J.S.:** Investigation. **C.P.:** Investigation. **M.F.:** Validation, Writing – review & editing. **Y.H.S.C.:** Investigation, Resources. **H.B.:** Resources, Validation. **R.S.:** Funding acquisition, Writing – review & editing. **P.S.:** Conceptualization, Funding acquisition, Methodology, Supervision, Writing – review & editing. **R.B.:** Supervision, Writing – review & editing. **J.A.:** Conceptualization (lead), Formal analysis (lead), Investigation, Methodology, Project administration (lead), Visualization, Writing – original draft, Writing – review & editing (lead).

All authors have read and approved the final version of the paper.

# 10. Funding

This work was supported by the Deutsche Forschungsgemeinschaft (DFG, German Research Foundation) under project 426208833. Additional DFG support was received through project 426206394 (Y.H.S.C., P.S., N.J.P. and R.S.) and project 426362670 (H.B., Y.H.S.C. and P.S.).

# 11. Competing Interests

The authors declare no competing interests.